\documentclass[a4paper,english]{iopart}
\usepackage[T1]{fontenc}
\usepackage[latin9]{inputenc}
\usepackage{textcomp}
\usepackage{amsbsy}
\usepackage{graphicx}

\usepackage{iopams}
\usepackage{setstack}

\usepackage{babel}

\begin{document}

\title{Post-Tanner spreading of nematic droplets}

\author{S Mechkov$^{1}$, A M Cazabat$^{2}$ and G Oshanin$^{1,3}$}

\address{$^{1}$Laboratoire de Physique Théorique de la Matière Condensée,
Université Pierre et Marie Curie, 4 place Jussieu, 75252 Paris Cedex
5 France}

\address{$^{2}$Laboratoire de Physique Statistique, Ecole Normale Supérieure,
75252 Paris Cedex 5 France}

\address{$^{3}$Laboratory J.-V. Poncelet (UMI CNRS 2615), Independent University
of Moscow, Bolshoy Vlasyevskiy Pereulok 11, 119002 Moscow, Russia}

\ead{mechkov@lptmc.jussieu.fr, anne-marie.cazabat@lps.ens.fr and oshanin@lptmc.jussieu.fr}
\begin{abstract}
The quasistationary spreading of a circular liquid drop on a solid
substrate typically obeys the so-called Tanner law, with the instantaneous
base radius $R(t)$ growing with time as $R\sim t^{1/10}$ -- an effect of the
dominant role of capillary forces for a small-sized droplet. However,
for droplets of nematic liquid crystals, a \emph{faster} spreading
law sets in at long times, so that $R\sim t^{\alpha}$ with $\alpha$
significantly larger than the Tanner exponent $1/10$. In the framework
of the thin film model (or lubrication approximation), we describe
this {}``acceleration'' as a transition to a qualitatively different
spreading regime driven by a strong substrate-liquid interaction specific
to nematics (antagonistic anchoring at the interfaces). The numerical
solution of the thin film equation agrees well with the available
experimental data for nematics, even though the non-Newtonian rheology
has yet to be taken into account. Thus we complement the theory of
spreading with a post-Tanner stage, noting that the spreading process can
be expected to cross over from the usual capillarity-dominated stage
to a regime where the whole reservoir becomes a diffusive film in
the sense of Derjaguin.
\end{abstract}

\vspace{2pc}
\maketitle

\section{Introduction}

\begin{figure}
\begin{centering}
\includegraphics[width=1\columnwidth]{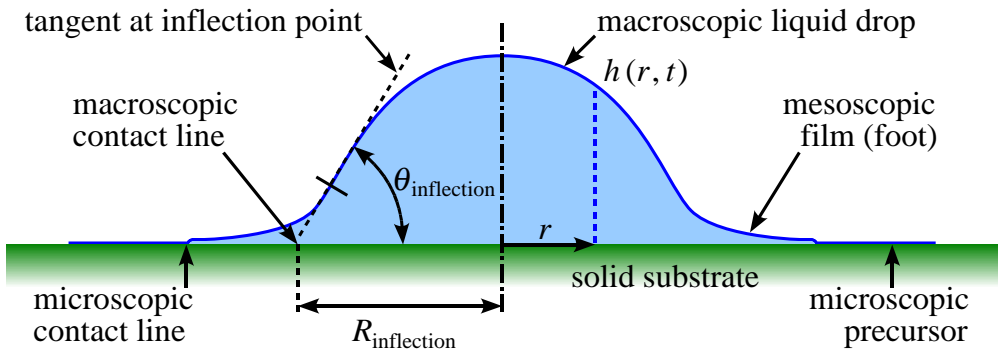} 
\par\end{centering}

\caption{\label{fig:DropCrossSection}Cross-section of a circular droplet spreading
in a situation of complete wetting (cartoon). See section \ref{sub:scales}
for details.}

\end{figure}

The spreading of liquid drops and films on a solid surface can be
described by universal, {}``macroscopic'' laws \cite{degennes85}
as soon as the thickness of the drop or film exceeds a few tens of
nanometers. One such law is the so-called Tanner law, characteristic
of the spontaneous spreading of small non-volatile drops on a flat substrate
in a situation of complete wetting (see figure \ref{fig:DropCrossSection}).
After an initial transient regime, the base radius $R$ of such a
drop grows as $R\sim t^{1/10}$. The law has been derived analytically
\cite{voinov,tanner,hervet} and confirmed experimentally on many
accounts \cite{tanner,aus,caz,we}. The fundamental argument is that
the hydrodynamics in the bulk of a drop are driven by capillary forces
alone, which directly yields $R\sim t^{1/10}$ assuming a self-similar
shape for the bulk, in the lubrication approximation \cite{voinov,tanner}.
Alternatively, the trend can be regarded as a competition between
the hydrodynamic dissipation (primarily in the contact line region
of the drop) and an unbalanced capillary force \cite{degennes85,hervet,cazabat,joanny}.

The Tanner law is quite robust and typically offers an accurate description
of the life of a droplet -- which spans a few hours for liquids with moderate 
surface tensions and viscosities -- apart from initial and final transients. 
The initial transient corresponds, e.g., to the deposit of the droplet on the
substrate, and lasts less than a second for regular liquids. As for the final 
state of spreading, for non-volatile droplets it is either a molecular film
or a flat, bounded structure -- a {}``mesoscopic'' pancake \cite{degennes85,ruck,joanny1984,joanny1984b}
-- which may be more favorable energetically than a molecular film.
Pancakes occur when short-range substrate interactions promote dewetting,
even though the overall situation is that of complete wetting \cite{derjaguin}:
although not very common, such structures have been observed experimentally
\cite{pancakes}. The existence of a limiting configuration, with
a finite value for the base radius $R$, implies that the late-time
spreading dynamics typically slow down with respect to the Tanner law.

By contrast to the trend of arrested spreading, it has recently been
observed that the Tanner stage $R\sim t^{1/10}$ can be followed by
a \textit{faster} spreading law $R\sim t^{\alpha}$, with $\alpha>1/10$.
Specifically, for spontaneously spreading nematic liquid crystals
\cite{cazb,cazc}, the value of $\alpha$ was found to be nearly twice
as large as the exponent $\alpha_{\mathrm{Tanner}}=0.1$ characterizing
the Tanner law, with $\alpha=0.2$ \cite{cazb} and $\alpha=0.19$
\cite{cazc}. A more thorough analysis of the data suggests that the
acceleration does not stop at $\alpha=0.2$: values as high as
$\alpha=0.3$ can be observed at the end of the experiment. This {}``accelerating''
trend is apparently in conflict with the notion of a Tanner stage
terminated by the onset of a molecular film or equilibrium pancake,
and its physical origin has yet to be clarified.

We have already attempted a qualitative explanation of this post-Tanner
trend in a macroscopic framework \cite{companion}. In the present
paper, our goal is to account for the acceleration quantitatively
and for this purpose we resort to the well-accepted thin film model
(TFM). Analyzing the thin film equation (TFE) we see that at late
spreading times the disjoining pressure dominates capillary effects.
Then the local thickness $h(r,t)$ (see figure 1) obeys a diffusion
equation, i.e., the whole droplet effectively becomes a diffusive
film in the sense of Derjaguin \cite{derjaguin}; the corresponding
base radius $R$ grows as $R\sim t^{1/2}$. The experimentally observed
transition from Tanner's law to power laws $R\sim t^{\alpha}$ with
$\alpha\approx 0.2$ \cite{cazb,cazc} in fact seems to be part of
a crossover to a much faster spreading law than expected previously.
In order to validate this observation we integrate the TFE numerically,
extract relevant observables and compare their evolution to the experimental
data. We find the general trend illustrated by the {}``numerical
spreading'' to be in good agreement with the spreading observed for
nematic droplets in \cite{cazb,cazc}. However, our model has yet
to take into account the typically non-Newtonian rheology of nematic
liquid crystals. Our work in progress will address this effect in separate publications. The macroscopic interpretation of the acceleration in terms 
of a negative line tension is also provided elsewhere \cite{companion}.

Our paper is organized as follows. In section 2 we give an overview
of our system of interest and summarize previously attempted explanations
for its abnormal spreading behavior. Section 3 features a brief derivation
of the thin film equation and presents key analytical results that
are directly relevant to the problem of nematic droplets. Section
4 focuses on key properties of our numerical spreading process (essentially
a brute-force solution of the TFE) , which is then compared quantitatively
with physical experimental data in section 5. We conclude in section
6, relating our results to alternative concepts and providing outlook
into our future work.

\section{Overview of the problem}

\subsection{Anatomy of a spreading droplet: macroscopic versus mesoscopic\label{sub:scales}}

Figure 1 represents the quasistationary state of a spreading droplet, 
which is composed of: a {}``macroscopic'' liquid drop where shear is
small and viscous forces are balanced primarily by variations of capillary
Laplace pressure; a {}``mesoscopic'' part subject to large shear,
where viscous forces are balanced primarily by variations of disjoining
pressure; a {}``microscopic'' region featuring molecular precursor
layers and a dry substrate \cite{ya,ya-1}. The relative sizes of
these regions are not up to scale: the main purpose of figure 1 is
to clearly distinguish the apparent contact line (typically inferred
from the inflection point of the profile $h(r,t)$ at a given time
$t$) from the {}``real'' contact line, governed by microscopic
phenomena.

Most analyses assume that the macroscopic and mesoscopic scales are
well-separated, i.e., that the bulk of the drop is much wider than
the mesoscopic {}``foot''. Thus the bulk is well-approximated
by a thin spherical cap with base radius $R(t)$, contact angle $\theta(t)\ll1$
and nearly constant volume \begin{equation}
V_{\mathrm{cap}}=\frac{\pi}{4}R^{3}\theta,\label{eq:Vcap}\end{equation}
i.e., it is practically in equilibrium at constant volume $V_{\mathrm{cap}}$
and instantaneous base radius $R(t)$. In this approximation it is
also customary to assimilate $R$ and $\theta$ with their respective
estimates inferred from the inflection point of $h(r,t)$, $R_{\mathrm{inflection}}$
and $\theta_{\mathrm{inflection}}$ (see figure 1), which can be obtained
through optical observation of the apparent contact line \cite{cazb,cazc}.
More immediately, these same optical experiments yield the location of the
microscopic contact line (the characteristic thickness of which is
30 nm) and, e.g., $R_{300\,\mathrm{nm}}(t)$ such that $h(R_{300\,\mathrm{nm}},t)=300\,\mathrm{nm}$
(see figures 2 and 4a). These measurements indicate that the length
of the mesoscopic film ahead of the apparent contact line grows to
millimetric sizes and becomes comparable with $R$ at the end of the
experiment. At this point the assumption of well-separated scales
is clearly broken, and the macroscopic volume $V_{\mathrm{cap}}$
is significantly lower than the total volume\begin{equation}
V=2\pi\int_{0}^{\infty} h\left(r,t\right)\,r\,\mathrm{d}r.\label{eq:Vtotal}\end{equation}

\subsection{Post-Tanner spreading laws: experimental evidence and tentative explanations}

\begin{figure}
\includegraphics[width=1\textwidth]{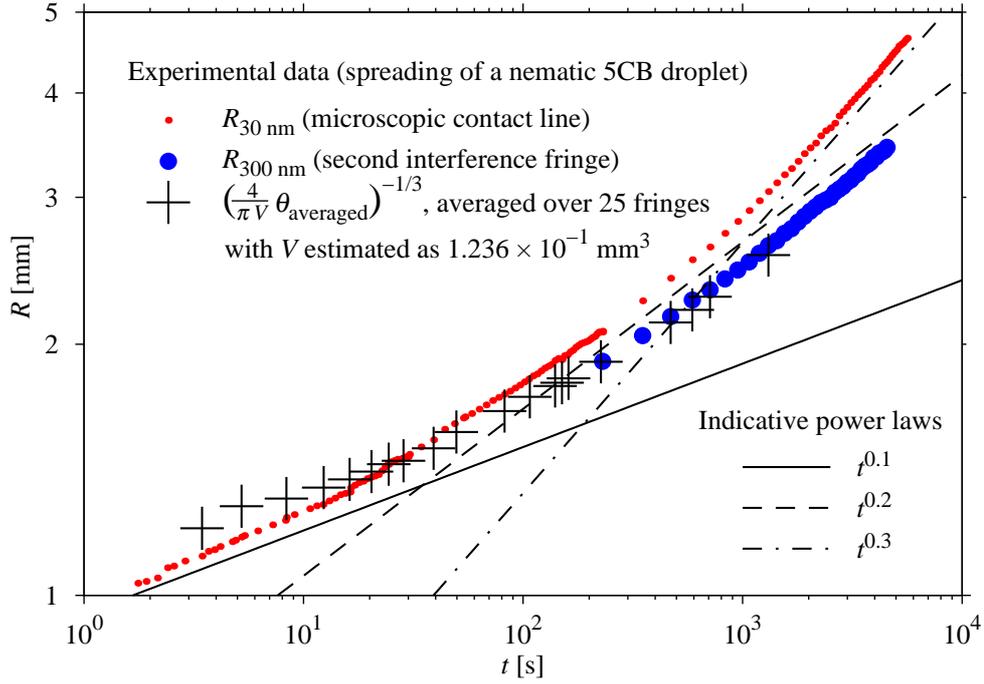}

\caption{Overview of the {}``accelerating'' trend for a single spreading
experiment (cyanobiphenyl 5CB droplet on silicon wafer). The quantities
observed are $R_{30\,\mathrm{nm}}$ (microscopic contact line; small
dots), $R_{300\,\mathrm{nm}}$ (second interference fringe; big
dots) and $\theta_{\mathrm{averaged}}$ (slope averaged over
25 interference fringes; crosses). At $t>200\,\mathrm{s}$ the plot
reveals the existence and growth of a large mesoscopic {}``foot''
at the edge of the droplet, comparable in size to the base radius
(see figures 1 and 4a). It is also clear that both observed radii
significantly deviate from Tanner's law $R\sim t^{1/10}$ and undergo
a transient {}``acceleration'', past both the $R\sim t^{1/5}$ and
$R\sim t^{3/10}$ power laws. The volume $V$ of the droplet was not
measured directly but it can be estimated as $V=1.236\,\times\,10^{-1}\,\mathrm{mm}^{3}$
(see sections \ref{sub:Scaling} and \ref{sec:Comparison}).}

\end{figure}
By contrast with previous evidence of Tanner's law, recent experimental
studies of spontaneous spreading of nematic liquid crystals on hydrophilic
\cite{cazb} or hydrophobic \cite{cazc} substrates revealed, after
a transient Tanner stage, a surprising {}``acceleration'' (actually
a spreading process that {}``slows down more slowly'' than the Tanner
law). Figure 2 plots the relevant observables for one spreading experiment.
The base radius $R$ and contact angle $\theta$ were reportedly inferred
from the inflection point of the thickness profile (see figure 1),
although in fact the main observables were $R_{30\,\mathrm{nm}}$
and $\theta_{\mathrm{averaged}}$ (see figure 4a) rather than $R_{\mathrm{inflection}}$
and $\theta_{\mathrm{inflection}}$. Initially \cite{cazb} it was
noted that Tanner's law crossed over to $R\sim t^{\alpha}$ with $\alpha\approx0.2$.
It was also realized (see figure 4b in \cite{cazb}) that the Tanner
relation $\theta^{3}\sim\mathrm{Ca}$ (where $\mathrm{Ca}\equiv\frac{\eta}{\sigma}\frac{\mathrm{d}R}{\mathrm{d}t}$
is the capillary number) does not hold for late spreading times:
for small $\theta$ and $\mathrm{Ca}$ one has $\theta\sim\mathrm{Ca}^{0.75}$.
The latter relation, together with the volume conservation condition
$R^{3}\theta\sim V$, is consistent with $R\sim t^{0.2}$. Similar
results were reported for spreading on hydrophobic substrates (see
figure 6 in \cite{cazc}), with $\theta\sim\mathrm{Ca}^{0.7}$ and
$R\sim t^{0.19}$.

The conclusions from these results are as follows. On one hand \cite{cazb},
direct estimates of $\alpha$ through $R(t)$ are consistent with
estimates via $\theta\left(\mathrm{Ca}\right)$, which apparently
validates the hypothesis of an approximating spherical cap of constant
volume (although the experimental data \cite{cazb} offers no \emph{direct} evidence for this). On the other hand, $\alpha$ is found to be 
significantly larger than the Tanner exponent $\alpha_{\mathrm{Tanner}}=0.1$. This signifies that some unknown
factor, other than the surface tension, comes into play. Moreover,
similar values of $\alpha$ were obtained for different kinds of substrates
\cite{cazb,cazc}, which suggests that the acceleration is a robust
effect rather than an artefact, and an intrinsic feature of nematic
droplets. One should also note that the experiment in \cite{cazc}
does not fully capture the post-Tanner transient and that exponents
as high as $\alpha\approx0.3$ are observed at late times (see figure
2).

Several qualitative arguments come to mind, which may or may not explain
the reported acceleration. First of all, nematic crystals are known
to have a non-Newtonian, shear-thinning rheology \cite{nakano2003,nakano2}.
Shear thinning affects the flow pattern and the dissipation rates
and thus modifies the spreading dynamics, but it is not clear 
a priori whether the actual dynamics will be faster or slower 
than Tanner's power laws.

A detailed analysis of the contact line dynamics in the framework
of the thin film model \cite{carre} shows that characteristic
shear rates in the capillary wedge and in the mesoscopic precursor
\emph{decrease} as the contact line slows down, and thus for a non-Newtonian
fluid the effective viscosity will \emph{increase} with time. This
corresponds to a modified spreading law $R\sim t^{\alpha}$ with $\alpha<1/10$.
Numerical experiments were carried out \cite{starov} and confirmed
$\alpha<1/10$ for shear-thinning fluids and $\alpha>1/10$ for shear-thickening
fluids. Thus the dominant effect from shear thinning is that
the spreading is \emph{slower} than predicted by Tanner's law, and
we must seek another mechanism to explain the {}``acceleration''
observed for 5CB droplets.

Among other factors that could be responsible for an acceleration of spreading, 
we should also cite: a) slippage at the substrate \cite{slip}; b) somewhat 
counterintuitively, densely distributed roughness, which for small or zero 
contact angles causes wicking and "enhances" the property of complete wetting 
\cite{mchale}. Unfortunately, we must discard both these effects as possible 
causes of the observed acceleration in the case of our nematic droplets. 
First, slip has been shown to cause a logarithmically small contribution to the macroscopic spreading laws \cite{slip}; the acceleration observed in 
\cite{cazb,cazc} looks qualitatively different from a minor effect due to slip. 
As for roughness, the nematic droplets may encounter some anchoring defects on hydrophilic substrates \cite{cazb}, but hydrophobic substrates in \cite{cazc} 
are definitely free of either chemical defects or topographic roughness; thus 
the consistent acceleration observed in both cases is not likely to be related 
to the "superwetting" properties of rough substrates.

Finally, a very tempting approach is to describe the (non volatile)
system in terms of its total free energy. The 1985 review by de Gennes
\cite{degennes85}
has explained Tanner's law in terms of an effective driving force
(derived from the instantaneous free energy). The work of the driving
force is balanced by dissipation, primarily hydrodynamic dissipation
in the macroscopic {}``wedge'' and mesoscopic {}``foot'' in the
vicinity of the apparent contact line. The assumptions of this approach
can be challenged by hypothesizing incomplete dissipation in the foot/precursor
\cite{cazb} or by introducing the concept of dynamic line tension,
which contributes to the unbalanced Young force and plays a dominant
role at long spreading times \cite{companion}.

We note, however, that line tension as an equilibrium concept is quite
subtle \cite{schimmele07,EPL} and its generalization to a quasistationary
situation should not be taken lightly. It is also hard to derive a
consistent set of correction terms for the hydrodynamic dissipation.
More generally, the notion of a driving force acting on the edge of
a macroscopic, capillary drop -- this notion breaks down when the
size of the mesoscopic region ({}``foot'') becomes comparable to
that of the bulk of the droplet, which is apparently the case during
the reported acceleration (this is indicated, e.g., by the evolution
of $R_{30\,\mathrm{nm}}$ and $R_{300\,\mathrm{nm}}$ on figure 2).
This prompts us to describe the spreading droplet in a framework that
resolves mesoscopic regions and does not use macroscopic approximations
-- the thin film model.

\section{Thin film equation: presentation and analytical results\label{sec:ThinFilmModel}}

The thin film model (TFM) -- related to both the {}``interface displacement
model'' and the {}``lubrication approximation'' -- is a continuum
representation of the spreading dynamics, suitable for the study of
thin films. While it may not accurately describe the spreading dynamics
at molecular film thicknesses (see section \ref{sub:cutoff}), it
is believed to work quite well for mesoscopic thicknesses, i.e., above
several tens of nanometers. As opposed to macroscopic frameworks,
the TFM accurately resolves quantities that would otherwise remain
empirical, e.g., functionals of the thickness profiles in the vicinity
of the apparent contact line and in the mesoscopic {}``foot'' of
a droplet. Notably, Tanner's law was consistently derived in the framework
of the TFM by Voinov \cite{voinov}, Tanner \cite{tanner} and de
Gennes \cite{degennes85,hervet}. Numerous authors have since used
the TFM to validate, refine, and generalize the features of advancing
contact lines and, by extension, the spreading dynamics of droplets
\cite{diez,gratton,eggers}.

\subsection{Thin film model as applied to nematic droplets}

At the core of the framework is the thin film equation (TFE). The
simplest expression of the TFE is for a Newtonian fluid with no slip
at the substrate, in the approximation of small thickness gradients.
In the following we briefly derive a TFE for nematic droplets.

We consider a quasistationary film of heterogeneous thickness $h\left(x,y\right)$
covering a homogeneous, flat substrate. Assuming that local equilibrium
is achieved for all $(x,y)$ and that the lateral flows in the film
have negligible inertia (low Reynolds number), we can write the following
energy functional:\begin{equation}
\mathcal{E}\left[h\right]=\int\int\left[\sigma+\sigma_{SL}+\frac{\sigma}{2}\left(\nabla h\right)^{2}+\Phi\left(h\right)\right]\mathrm{d}x\,\mathrm{d}y.\label{eq:EnergyFunctional}\end{equation}
Here $\sigma$ and $\sigma_{SL}$ are the nominal surface energies
of a free interface and of a solid-liquid interface, respectively,
whereas $\Phi\left(h\right)$ is an effective interface potential
acting as a correction to $\sigma+\sigma_{SL}$ due to the fact that
$h$ is finite. As for the excess energy due to the curvature of the
free interface, it is well approximated with $\frac{\sigma}{2}\int\int\left(\nabla h\right)^{2}\,\mathrm{d}x\,\mathrm{d}y$
in the small-slope approximation $\left|\nabla h\right|\ll1$. It is common to neglect hydrostatic contributions to (\ref{eq:EnergyFunctional}) in situations 
of complete wetting \cite{degennes85, eggers}.

Considering $\mathcal{E}\left[h\right]$ under volume-preserving variations
of $h$, the quasistationary internal pressure $p\left(x,y\right)$
is found to be of the intuitive form \begin{equation}
p=-\sigma\Delta h-\Pi\left(h\right),\label{eq:QuasistationaryPressure}\end{equation}
which is a combination of the typical {}``capillary'' Laplace pressure
and of the {}``disjoining'' pressure $\Pi\left(h\right)\equiv-\partial_{h}\Phi\left(h\right)$.
The lateral pressure gradient $\nabla p$ is relaxed through a so-called
Poiseuille flow \begin{equation}
\mathbf{j} = -\frac{h^{3}}{3\eta}\nabla p,
\label{eq:PoiseuilleCurrent}\end{equation}
assuming a constant viscosity $\eta$ and no slip at the substrate 
(for a derivation, see \cite{degennes85} or \cite{eggers}). Finally,
the conservation equation $\partial_{t}h=-\nabla\cdot\mathbf{j}$,
together with (\ref{eq:QuasistationaryPressure}) and 
(\ref{eq:PoiseuilleCurrent}),
yields\begin{equation}
\partial_{t}h=-\nabla\cdot\left\{ \underbrace{-\frac{h^{3}}{3\eta}\nabla\left[\underbrace{-\sigma\Delta h-\Pi\left(h\right)}_{\mbox{pressure }p}\right]}_{\mbox{lateral current }\mathbf{j}}\right\} .\label{eq:TFE0}\end{equation}
Due to the different nature of the two contributions to the pressure
$p$, it is appropriate to rewrite (\ref{eq:TFE0}) as\begin{equation}
\partial_{t}h=-\frac{\sigma}{\eta}\nabla\cdot\left(\frac{1}{3}h^{3}\nabla\Delta h\right)+\nabla\cdot\left[D(h)\nabla h\right]\label{eq:TFE2}\end{equation}
where \begin{equation}
D=-\frac{h^{3}}{3\eta}\frac{\mathrm{d}\Pi}{\mathrm{d}h}\label{eq:Dh}\end{equation}
is the effective diffusion coefficient introduced by Derjaguin \cite{derjaguin}.
The expression of $D$ contributes to the second-order term of (\ref{eq:TFE2})
and plays a major role in the spreading dynamics at mesoscopic
thicknesses. In a continuum representation, it is expedient to approximate
the disjoining pressure $\Pi(h)$, and hence $D(h)$, with a dominant
long-range contribution, while introducing a phenomenological boundary
condition (effectively a cutoff) in the nanometric range. Previous
studies \cite{degennes85,hervet,eggers} addressed the case of $\Pi\left(h\right)=\frac{1}{6\pi}A\, h^{-3}$,
a single power law accounting for the cumulated effect of non-retarded
van der Waals interactions for a film of finite thickness $h$ ($A$
being the Hamaker constant). However, in the case of antagonistically
anchored nematic liquid crystals \cite{cazb,cazc}, the dominant term
is \begin{equation}
\Pi=\frac{1}{2}K\delta^{2}h^{-2},\label{eq:DisjoiningPressureElastic}\end{equation}
$K$ being the bend-splay elastic constant and $\delta$ the angle
by which the director rotates over the thickness of the film, i.e.,
the difference between the anchoring angles at both interfaces. Note
that we only take into account the elastic energy in the bulk of the
nematic; the anchoring energies (surface terms) are
taken to be constant, i.e., we assume sufficiently strong anchoring
at both interfaces with respect to the thickness of the film. Thus
in our case (\ref{eq:Dh}) reduces to\begin{equation}
D=\frac{K\delta^{2}}{3\eta}.\label{eq:D}\end{equation}
Interestingly, the elastic interaction typical of antagonistically
anchored nematics yields a purely diffusive film in the sense of Derjaguin.
To the best of our knowledge, this remarkable feature of a thickness-independent
diffusion coefficient has not been emphasized previously.

When describing the spreading droplet as a whole (as opposed to assimilating
the apparent contact line to a quasistationary hydrodynamic wedge
\cite{degennes85,hervet,eggers}), it is appropriate to rewrite (\ref{eq:TFE2})
in a \emph{rotationally} invariant geometry, i.e., with $h$ depending
only on the distance $r$ to the vertical axis of the droplet (see
figure 1), \begin{equation}
\partial_{t}h=-\frac{\sigma}{\eta}\frac{1}{r}\partial_{r}\left\{ \frac{1}{3}h^{3}r\partial_{r}\left[\frac{1}{r}\partial_{r}\left(r\partial_{r}h\right)\right]\right\} +D\frac{1}{r}\partial_{r}\left(r\partial_{r}h\right).\label{eq:TFE1DR}\end{equation}
Here we have used the fact that $D$ is a constant for nematics.

\subsection{Asymptotic spreading behavior}

Before we use (\ref{eq:TFE1DR}) for quantitative predictions, we
can make an important qualitative remark about the two limiting cases of 
the TFE. Looking at the two terms on the right-hand side of (\ref{eq:TFE2}) 
or (\ref{eq:TFE1DR}), for sufficiently tall droplets, the disjoining 
pressure $\Pi=\frac{1}{2}K\delta^{2}h^{-2}$ is negligible with respect
to the Laplace pressure over a large part of the droplet (the droplet
is well-approximated by a spherical cap as in section \ref{sub:scales}
and the Laplace pressure is $2\sigma\theta/R\simeq2\pi\left[h(r=0)\right]^{2}/V_{\mathrm{cap}}$).
In this approximation, the TFE is essentially fourth-order and has
the form\begin{equation}
\frac{3\eta}{\sigma}\partial_{t}h=-\frac{1}{r}\partial_{r}\left\{ h^{3}r\partial_{r}\left[\frac{1}{r}\partial_{r}\left(r\partial_{r}h\right)\right]\right\} .\label{eq:TFEtanner}\end{equation}
Looking for self-similar solutions of the form $h\left(r,t\right)=t^{-2\alpha}f\left(t^{-\alpha}r\right)$,
with the scaling chosen so that $V=2\pi\int h(r,t)\, r\,\mathrm{d}r$
remains constant, (\ref{eq:TFEtanner}) yields $\alpha=1/10$, i.e.,
Tanner's law \cite{voinov,tanner,diez,gratton}.

It is also clear that at late stages of spreading, as the droplet
becomes flatter, the Laplace pressure will eventually be dominated
by the disjoining pressure, and the TFE {[}cf. (\ref{eq:TFE2}) and
(\ref{eq:TFE1DR}){]} will be essentially second-order, governed by
the specific liquid-substrate interactions:\begin{equation}
\partial_{t}h=\frac{1}{r}\partial_{r}\left[r\, D(h)\partial_{r}h\right].\label{eq:TFEdiffusion}\end{equation}
In the nematic case, $D$ has the constant expression (\ref{eq:D})
and an obvious self-similar solution of (\ref{eq:TFEdiffusion}) is
a Gaussian bell defined by \begin{equation}
h(r,t)=\frac{V}{4\pi Dt}\exp\left(-\frac{r^{2}}{4Dt}\right),\label{eq:GaussianBell}\end{equation}
with an arbitrary origin for time. The base radius and contact angle
as inferred from the inflection point are $R=\sqrt{8Dt}$ and $\theta=\frac{4V}{\pi\sqrt{\mathrm{e}}}R^{-3}$,
so that the volume conservation relationship is $V=\sqrt{\mathrm{e}}\frac{\pi}{4}R^{3}\theta$,
as opposed to $V=\frac{\pi}{4}R^{3}\theta$~for the idealized Tanner
regime.

Thus we see that the spreading process must cross over from an initial
spreading phase, consistent with the generic Tanner's law, to another
regime, specific to antagonistically anchored nematic liquid crystals:
we shall see in section \ref{sub:Scaling} that the characteristic
time of the crossover scales as $T=\left(\frac{\sigma}{\eta}V^{3}/D^{5}\right)^{1/4}$.
The late-time evolution of a droplet is expected to be \emph{diffusive},
i.e., measurements of $R(t)$ will yield an {}``acceleration'' from
$R\sim t^{1/10}$ to $R\sim t^{1/2}$. The remaining problem is to
establish the characteristics of the crossover and to compare it to
the physical experiment.

\section{Numerical integration of the TFE: preliminaries}

\begin{figure}
\includegraphics[width=1\columnwidth]{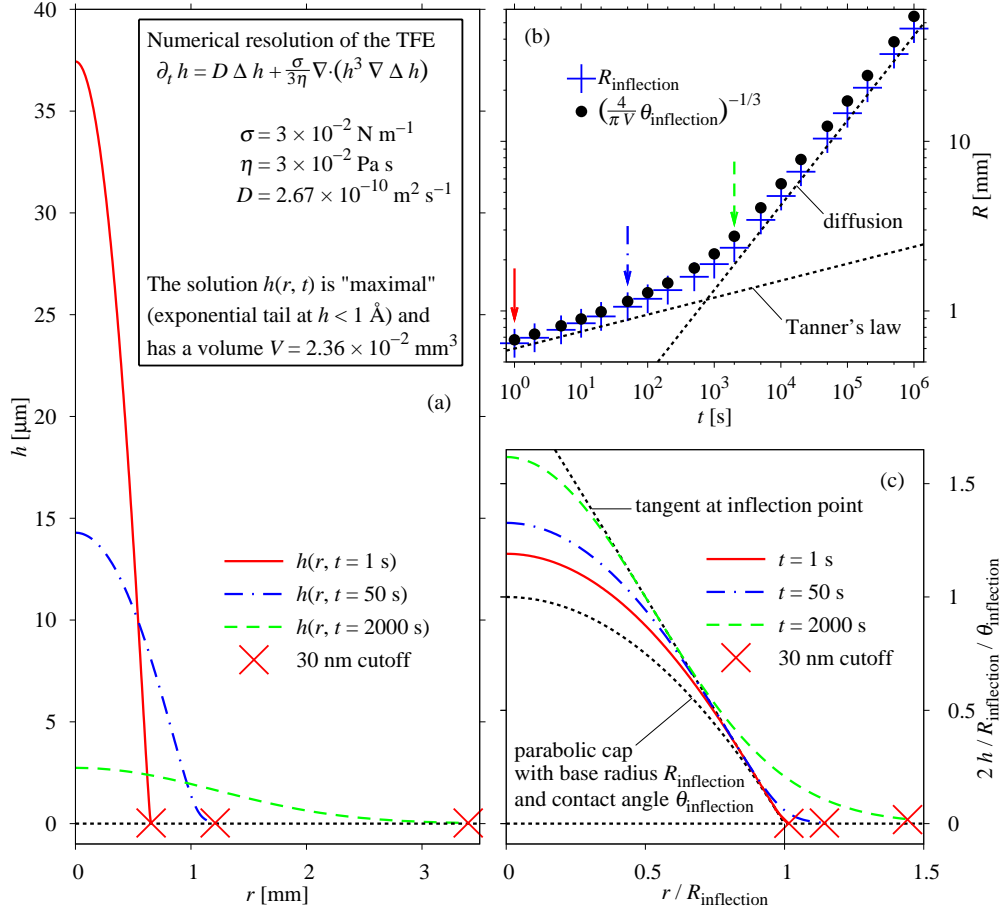}

\caption{\label{fig:numerical}Overview of the results of the numerical resolution
of the thin film equation (TFE) with an {}``elastic'' contribution
to the disjoining pressure, i.e., a diffusive second-order term: a)
three consecutive shapes adopted by a droplet of volume $V=2.36\,\times\,10^{-2}\,\mathrm{mm}^{3}$
during a numerical spreading process (the oblique crosses indicate
the points at which the solution $h\left(r,t\right)$ falls below
$\varepsilon=30\,\mathrm{nm}$, which is the characteristic thickness
at the edge of the physical mesoscopic precursor \cite{cazc}); b)
same experiment: time dependence of the base radius $R$ of the drop,
as estimated directly from the tangent at the inflection point (see
figure 1) and also, tentatively, from the volume conservation law
$V=\frac{\pi}{4}R^{3}\theta$; c) comparative analysis of the estimates
of $R$ and $\theta$ for the same cross-section snapshots as in (a):
the tangent at the inflection point yields a sensible value for $R$,
but $\theta$ does \emph{not} correspond to an apex-fitting spherical
cap, even at $t=1\,\mathrm{s}$.}

\end{figure}

We integrated (\ref{eq:TFE1DR}) in the form of a numerical spreading
process, taking snapshots of the solution $h(r,t)$ at preset time
intervals (figure \ref{fig:numerical}a). The numerical values
used were $\sigma=30\,\times\,10^{-3}\,\mathrm{N}\cdot\mathrm{m}^{-1}$, $\eta=30\,\times\,10^{-3}\,\mathrm{Pa}\cdot\mathrm{s}$
and $K\delta^{2}=12\,\times\,10^{-12}\,\mathrm{J}\cdot\mathrm{m}^{-1}$ (i.e., $D=2.67\,\times\,10^{-10}\,\mathrm{m}^{2}\cdot\mathrm{s}^{-1}$).
As for the boundary conditions, below $h=1$\AA~we extrapolate
$h(r,t)$ as an exponential tail (similar to a {}``maximal film''
\cite{degennes85,hervet,eggers}) and ensure that the total volume
$V=2\pi\int h(r,t)\, r\,\mathrm{d}r$ is preserved: in figure 
\ref{fig:numerical}, $V=2.36\,\times\,10^{-2}\,\mathrm{mm}^{3}$. 
As for the initial conditions,
we start with a perfect parabolic cap, to which we add a moderately
smooth foot to avoid a computationally heavy singularity. Integration
is explicit in time, with a fixed-size grid for $r$ and an adaptive
time step.%

Figure \ref{fig:numerical}b plots the spreading versus time in terms
of the base radius $R$ and contact angle $\theta$, as inferred from
the inflection point of $h(r,t)$ (see figure 1 and figure \ref{fig:numerical}c).
The volume of the numerical droplet is similar to the physical experiment
\cite{cazc}, as well as the characteristics 
$\sigma=30\,\times\,10^{-3}\,\mathrm{N}\cdot\mathrm{m}^{-1}$,
$\eta=30\,\times\,10^{-3}\,\mathrm{Pa}\cdot\mathrm{s}$ and $K\delta^{2}=12\,10^{-12}\,\mathrm{J}\cdot\mathrm{m}^{-1}$.
The acceleration agrees qualitatively with figure 2 (cf. figure 5
in \cite{cazc}): the crossover from the Tanner phase to significantly
faster regimes occurs at times of the order of a minute, for a droplet
of the same {}``caliber'' as in \cite{cazc} ($R\simeq1\,\mathrm{mm}$
at $t=1\,\mathrm{s}$); a characteristic time of the crossover corresponds
to the intersection of the asymptotes in figure 3b, at $t\simeq700\,\mathrm{s}$
(about 12 minutes).

Before comparing quantitatively the physical spreading experiment
with our numerical resolution of the TFE, we shall voice a few words
of caution about the applicability of the TFM. We shall also review
the physical observables available to us. Finally, we shall discuss
the issue of time origin in spreading experiments and the scaling
feature of equation (\ref{eq:TFE1DR}).

\subsection{TFM applicability\label{sub:cutoff}}

The oblique crosses on figures \ref{fig:numerical}a and \ref{fig:numerical}c
indicate the points at which the solution $h\left(r,t\right)$ falls
below $\varepsilon=30\,\mathrm{nm}$, which is the characteristic
thickness at the edge of the physical mesoscopic precursor \cite{cazc}.
For $t=2000$ s (dash-dotted profile), $h(r=0)\simeq100\varepsilon$,
thus the maximal solution $h(r,t=2000\,\mathrm{s})$, as a continuum
construct, is at the limit of physical relevance. We also note that
the experimentally observed spreading dynamics of a similar 5CB droplet
typically stop after an hour, with $R$ of the order of tens of millimeters
(which corresponds to a pancake of volume $V\simeq10^{-2}\,\mathrm{mm}^{3}$
and thickness 30 nm). This behavior is due to short-range interactions
which promote dewetting, and can not be captured by the TFE (\ref{eq:TFE1DR})
unless the expression (\ref{eq:DisjoiningPressureElastic}) is refined.
However, we \emph{do} expect the TFE to capture the relevant properties
of the contact line region as observed experimentally by Cazabat et
al \cite{cazb,cazc}, provided that the $30\,\mathrm{nm}$ thickness
plays the role of a cutoff, located in the asymptotic region of $h(r,t)$
as is still the case for $t=2000\,\mathrm{s}$ on figures \ref{fig:numerical}a
and \ref{fig:numerical}c.

\subsection{Review of the observables}

Figure \ref{fig:numerical}c makes it clear that the crossover from
Tanner's law to a faster regime (at times of the order of a minute
to an hour) coincides with a gradually lesser separation between the
macroscopic and mesoscopic scales. In other terms, the extent of the
mesoscopic {}``foot'' of the drop as compared to the macroscopic
{}``cap'' is such that the apparent contact line is ill-defined.
Another, somewhat unexpected fact shown by figure \ref{fig:numerical}c
is that measuring the contact angle $\theta_{\mathrm{inflection}}(t)$
at the (mesoscopic) inflection point of $h(r,t)$ does not give a good
estimate of $\theta$ for the (macroscopic) apex-fitting spherical
cap, even in the Tanner phase (at $t=1\,\mathrm{s}$) and despite
a seemingly good separation of the scales. This observation, however,
does not challenge our study, provided that we complement $R$, $\theta$,
$R_{\mathrm{inflection}}$ and $\theta_{\mathrm{inflection}}$ with
additional observables that are consistent with those measured by
Cazabat et al \cite{cazb,cazc}.

As a matter of fact, the optical measurements in \cite{cazb,cazc}
do not infer $R$ and $\theta$ from the inflection point, which would
have required a thorough reconstruction of the profile at the contact
line for each snapshot. It was more expedient to track the edge of
the spreading drop (located at an approximate thickness of 30 nm;
we shall note this radius $R_{30\,\mathrm{nm}}$) or the second interference
fringe of the ordinary-extraordinary coincidence pattern (at a thickness
of about 300 nm; we shall note this thickness $R_{300\,\mathrm{nm}}$).
As for the contact angle, the slope at the contact line was averaged
over the first 25 interfringes of the same pattern (between 300 nm
and 5.3 \textmu{}m). The numerical counterparts to these observable
characteristics are presented in figure \ref{fig:accel-fit}a, on
a representative snapshot of the numerical spreading process. We note
that the interference pattern used here is specific to nematics: the
fringes correspond to coincidence between the ordinary and extraordinary
rays, and the interfringe is about 11 times larger than for the normal
equal-thickness fringes.

\subsection{Time origin of spreading processes}

Both physical and numerical spreading processes can be seen as subject
to initial conditions such as the deposit of a drop. Typically, shortly
after a sufficiently compact deposit, flow patterns appear at the
edge of the drop and propagate throughout the initially static droplet,
establishing Tanner's regime; at later times, the spreading crosses
over to, e.g., a diffusive phase (for nematic droplets with antagonistic
anchoring, as considered in this paper; c.f. figure \ref{fig:numerical}b).

The physically relevant deposit is closely related to the subtle issue
of choosing a time origin ($t=0$). On one hand, the exact history
of the deposit has no effect at the scale of the long spreading process.
On the other hand, the power-law behavior typically observed in spreading
is best represented in log-log diagrams (figure 2, 3b, 3c, 4b, 4c),
which are quite sensitive to the origin of $t$ at small spreading
times.

Thanks to the robust presence of a Tanner stage at early spreading
times, the dilemma is customarily resolved by describing the spreading
in terms of the time elapsed since the effective origin of the Tanner
phase: practically, for a sufficiently compact deposit, the origin
of $t$ is slightly adjusted so that, e.g., $R(t)$ is well-fit by
a $t^{1/10}$ power law at early spreading times. This may seem arbitrary
but is in fact fundamental in the sense that for increasingly compact
deposits of a given volume $V$ the spreading processes converge towards
a well-defined limiting process, which (at least in the framework
of thin film dynamics) precisely corresponds to a backwards extrapolation
of Tanner's law. This convention is adopted for both the physical
and numerical spreading processes presented in this paper.

\subsection{Scaling\label{sub:Scaling}}

A prominent feature of the model TFE (\ref{eq:TFE1DR}) is that the
equation can be scaled in terms of $h$, $r$ and $t$ (which corresponds
to three degrees of freedom on $\sigma$, $\eta$, $D$ and $V$).
If we know a function $h_{0}(r,t)$ that is a solution of \begin{equation}
\partial_{t}h_{0}=-\frac{\sigma_{0}}{\eta_{0}}\frac{1}{r}\partial_{r}\left\{ \frac{1}{3}h_{0}^{3}r\partial_{r}\left[\frac{1}{r}\partial_{r}\left(r\partial_{r}h_{0}\right)\right]\right\} +D_{0}\frac{1}{r}\partial_{r}\left(r\partial_{r}h_{0}\right)\label{eq:TFE1DRbis}\end{equation}
bearing the volume $V_{0}=2\pi\int_{0}^{\infty}h\left(r,t\right)r\mathrm{d}r$,
then we can define \begin{eqnarray}
k & \equiv & \left(V/V_{0}\right)^{1/8}\label{eq:kkk}\\
m & \equiv & \left(\frac{\sigma_{0}\eta D}{\sigma\eta_{0}D_{0}}\right)^{1/8}\label{eq:mmm}\\
n & \equiv & \left(\frac{\sigma_{0}\eta D^{5}}{\sigma\eta_{0}D_{0}^{5}}\right)^{1/4}\label{eq:nnn}\end{eqnarray}
and obtain a similar function $h(r,t)=k^{2}m^{2}h_{0}(k^{-3}m\, r,k^{-6}n\, t)$
which is a solution of (\ref{eq:TFE1DR}) with volume $V$.

In the work presented here, besides the obvious fitting in terms of
the volume $V$ via $k=\left(V/V_{0}\right)^{1/8}$, we assumed that
the surface tension $\sigma$ and the elastic coefficient $K\delta^{2}$
were not significantly different from the values 
$\sigma=30\,\times\,10^{-3}\,\mathrm{N}\cdot\mathrm{m}^{-1}$ and 
$K\delta^{2}=12\,\times\,10^{-12}\,\mathrm{J}\cdot\mathrm{m}^{-1}$. We allowed,
however, for an adjustment in terms of the effective viscosity $\eta$,
whereby $m=1$ and $n=\eta_{0}/\eta$. Indeed, the rheology of a nematic
film with antagonistic anchoring conditions is not as trivial as the
Poiseuille flow in our model TFE: the effective viscosity must be
intermediate between that of flow-aligned 5CB molecules ($30\,\times\,10^{-3}\,\mathrm{Pa}\cdot\mathrm{s}$)
and that of flow-orthogonal molecules ($100\,\times\,10^{-3}\,\mathrm{Pa}\cdot\mathrm{s}$)
\cite{nakano2003,nakano2}. This also affects the value of the effective
diffusion coefficient $D=\frac{K\delta^{2}}{3\eta}$. The results
of the fit are $\eta=70.5\,\times\,10^{-3}\,\mathrm{Pa}\cdot\mathrm{s}$, $D=1.136\,\times\,10^{-10}\,\mathrm{m}^{2}\cdot\mathrm{s}^{-1}$
and $V=1.236\,\times\,10^{-1}\,\mathrm{mm}^{3}$.

From the scaling factors $k$ and $n$ we conclude that the {}``characteristic
time'' of the crossover for the TFE (\ref{eq:TFE1DR}) scales as
$T=\left(\frac{\sigma}{\eta}V^{3}/D^{5}\right)^{1/4}$. For the numerical
values yielded by the fit we have $T\simeq8\,\times\,10^{4}\,\mathrm{s}$
(about 22 hours), which exceeds by far the duration of the physical
experiment (2 hours). From the intersection of the asymptotes on figure
3b (at $t\simeq700\,\mathrm{s}$) and the values $k=1.23$ and $n=0.425$
we can extract a more quantitatively relevant time $T=5.7\,\times\,10^{3}\,\mathrm{s}$,
i.e., a couple of hours rather than a day. Both values are consistent
with the fact that the crossover to a diffusive spreading process
is far from complete at the end of the observation in \cite{cazc}.

\section{Comparison of physical and numerical spreading processes\label{sec:Comparison}}

\begin{figure}
\includegraphics[width=1\columnwidth]{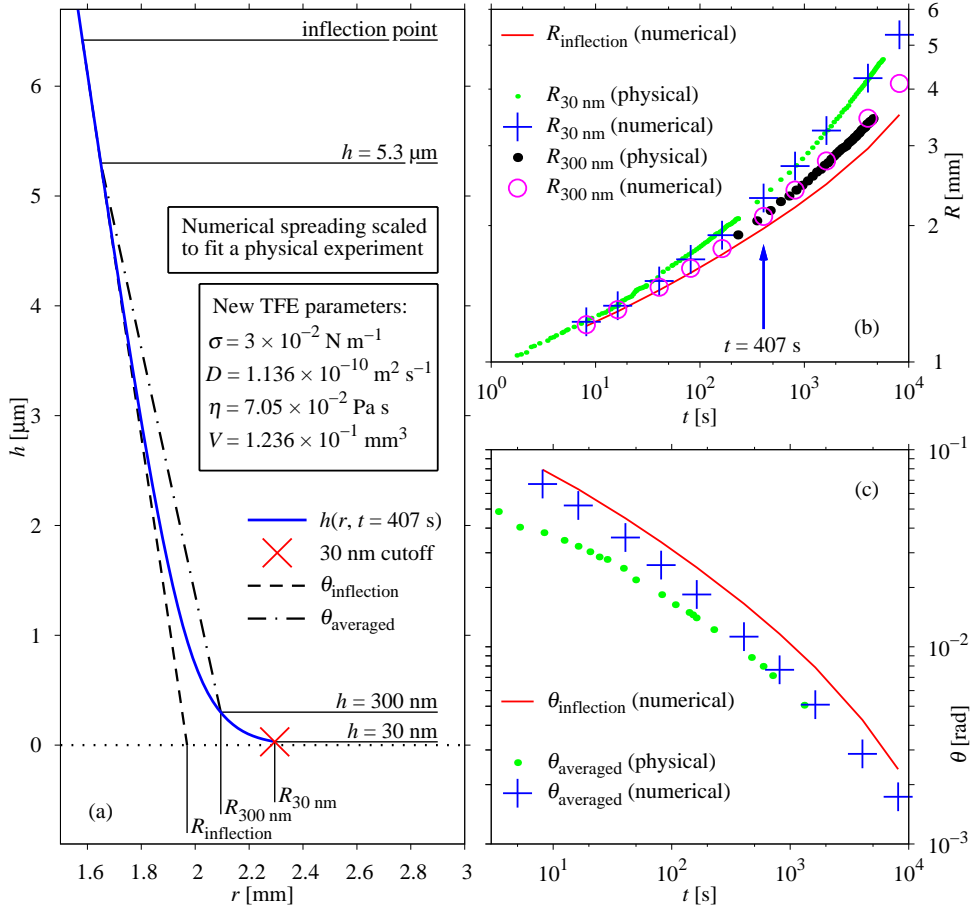}

\caption{\label{fig:accel-fit}Quantitative fit of a physical spreading experiment
(cyanobiphenyl 5CB on silicon wafer) in terms of the numerical spreading
process presented in figure 3: the volume $V$ and viscosity $\eta$
act as scaling parameters of the maximal solution obtained previously
(see text). Figure (a) illustrates the optically measured radii $R_{30\,\mathrm{nm}}$
(at the edge of the mesoscopic precursor) and $R_{300\,\mathrm{nm}}$
(corresponding to the second of a series of interference fringes),
as well as the optically measured estimate of the contact angle $\theta_{\mathrm{averaged}}$
(averaged over 25 fringes). Clearly $R_{\mathrm{inflection}}<R_{300\,\mathrm{nm}}<R_{30\,\mathrm{nm}}$
and $\theta_{\mathrm{inflection}}>\theta_{\mathrm{averaged}}$. Figure
(b) illustrates the good agreement for $R_{30\,\mathrm{nm}}$ and
$R_{300\,\mathrm{nm}}$. Figure (c) compares estimates of the contact
angle, with a satisfactory agreement at late spreading times, and
a discrepancy at early spreading times.}

\end{figure}

We shall now perform a quantitative matching between the physical
and numerical spreading processes. We adjusted the scaling of the
numerical solution to accommodate the physical experiment, shifting
the volume to $V=1.236\,\times\,10^{-1}\,\mathrm{mm}^{3}$ and the viscosity
to $\eta=7.05\,\times\,10^{-2}\,\mathrm{Pa}\cdot\mathrm{s}$. Then, for this
rescaled numerical experiment, we measured the same quantities as
observed optically, namely $R_{30\,\mathrm{nm}}$, $R_{300\,\mathrm{nm}}$
and $\theta_{\mathrm{averaged}}$ (see figure \ref{fig:accel-fit}a).
The results are presented in figures \ref{fig:accel-fit}b and \ref{fig:accel-fit}c.%

The best agreement was obtained for $R_{30\,\mathrm{nm}}$, over the
whole range where the two experiments overlap. As for the measurements
based on interference patterns, the agreement is less consistent.
On one hand (figure \ref{fig:accel-fit}b), the radius $R_{300\,\mathrm{nm}}$
of the second interference fringe agrees well with its numerical estimate.
On the other hand (figure 3c), at small spreading times the numerical
$\theta_{\mathrm{average}}$ is in excess of the optically observed
slope averaged over the first 25 interfringes. The latter discrepancy
may be due either to the low resolution of the fringes at early spreading
times or to our failure to capture complex shear-thinning effects
in the framework of the TFM.

Although the agreement is quite satisfactory for the $R_{30\,\mathrm{nm}}$
observable, we must note that the numerical estimate of $R_{30\,\mathrm{nm}}$
is at the limit of applicability of the TFM. In our model we simply
cut off the maximal solution of the TFE at the thickness $h=30$ nm,
and the dynamics of this cutoff line may differ from the actual dynamics
of the mesoscopic precursor near the microscopic contact line. It
would be more conclusive if the physical experiment had systematically
provided the more robust observables $R_{\mathrm{inflection}}$ and
$\theta_{\mathrm{inflection}}$.

\section{Conclusion}

We have attempted an explanation of the abnormal spreading properties
observed for small droplets of 5CB nematic liquid crystals \cite{cazb,cazc}
in the framework of the thin film model. This approach enabled
us to illustrate both qualitatively and quantitatively the key trends
in the spreading of nematic droplets:
\begin{itemize}
\item the development of a large {}``foot'' (mesoscopic precursor), whereby
the macroscopic and mesoscopic length scales are no longer well-separated;
\item the transition towards a faster spreading regime -- determined by the 
antagonistic anchoring of the nematic at the interfaces -- in which the 
thickness profile is essentially governed by a diffusion equation.
\end{itemize}
The {}``acceleration'', initially observed optically by Cazabat
et al \cite{cazb,cazc}, was reproduced in a numerical spreading process
(figure 3), which was used to fit the optical data (figure 2). The
primary optical observable being the edge of the mesoscopic precursor,
the agreement is satisfactory (figure 4b).

We note that this post-Tanner regime is a priori not specific to nematic
droplets. Similar crossovers to a faster spreading law than $R\sim t^{1/10}$
may be observed for regular liquids dominated by van der Waals forces,
although perhaps not as readily as in the present case, where the spreading
is driven by nematic elasticity. The fundamental result is that, for
long-range substrate interactions, the droplet essentially becomes
a diffusive film in the sense of Derjaguin \cite{derjaguin} at late
spreading times, and Tanner's law is gradually replaced with another
law, determined by the substrate interaction rather than by capillarity.

The work presented is a necessary complement to the quasistationary
energetic approach as presented in \cite{companion}, where acceleration
is interpreted in terms of a dynamic, negative line tension $\tau$
attributed to the apparent contact line (see figure 1). The latter
framework is applicable if the macroscopic and mesoscopic length scales
remain well-separated, i.e., if the bulk of the droplet is well-approximated
by a spherical cap, and if both the vertical and lateral size of the
mesoscopic region remain negligible. In this case, it is possible
to isolate a line contribution, which resides in the mesoscopic {}``foot'',
yet contributes to spreading dynamics at the macroscopic scale. However,
as the droplet spreads, it eventually adopts a characteristic bell
shape, and the capillarity-dominated spherical cap ceases to be a
good approximation. At even later stages, the droplet may reach the
state of a mesoscopic pancake, which can not be resolved by the macroscopic
model at all, unless the dimension is lowered to a planar geometry.

By contrast, the thin film model appears to provide a more robust description
of complete wetting situations, especially in the late stages of spreading.
In order to account for the emergence of pancakes, we aim to provide
the thin film equation with suitable boundary conditions that would
account for the phenomenology of the microscopic contact line. As
a prospect of future work, we also note that our current implementation of 
the thin film model does not accurately describe the non-Newtonian rheology
of antagonistically anchored nematics. In future studies we may refine
the notion of effective viscosity and allow for a more accurate modeling
of the spreading dynamics.

\ack{The authors gratefully acknowledge helpful discussions with H Tanaka,
S Dietrich and S V Meshkov. The experimental data displayed in figures
2, 4b and 4c was kindly provided by C Poulard.}

\end{document}